
\magnification=1200
\input harvmac 
\baselineskip=11pt
\parskip 5pt 
\hfuzz=18pt 

\def\np{\vfil\eject} 
\def\nl{\hfil\break} 
\def\nt{\noindent} 
\def\({\left(}
\def\){\right)}
\font\male=cmr9 
 
\font\fmale=cmr8 
\font\sfont=cmbx10 scaled\magstep1

\global\newcount\subsecno \global\subsecno=0
\global\newcount\meqno \global\meqno=1
\global\newcount\subsubsecno \global\subsubsecno=0

\def\newsubsec#1{\global\advance\subsecno
by1\message{(\the\secno.\the\subsecno. #1)}
\global\subsubsecno=0 

\noindent{\bf\the\secno.\the\subsecno. #1}\writetoca{\string\quad
{\the\secno.\the\subsecno.} {#1}}
\par\nobreak\medskip\nobreak}

\def\newsubsec#1{\global\advance\subsecno by1\message{(\secsym\the\subsecno.
#1)} \ifnum\lastpenalty>9000\else\bigbreak\fi 
\noindent{\bf\secsym\the\subsecno. #1}\writetoca{\string\quad 
{\secsym\the\subsecno.} {#1}}}

\def\newsubsubsec#1{\global\advance\subsubsecno
by1\message{(\the\secno.\the\subsecno.\the\subsubsecno.
#1)} \ifnum\lastpenalty>9000\else\bigbreak\fi 
\noindent{\bf\the\secno.\the\subsecno.\the\subsubsecno.
#1}\writetoca{\string\quad
{\the\secno.\the\subsecno.\the\subsubsecno.}
{#1}}}


\def\bu{$\bullet$} 
\def\hm{{\hat m}} \def\ha{{\hat a}}
\def\id{{\bf 1}} 
\def\va{{\vert a\vert}} \def\vv{{\vert a'\vert}}

\def\td{{\tilde d}} 
\def\had{{\textstyle{d\over2}}} 
\def\tdd{{[\had]}}

\def\thd{{\hat d}}
\def\hadd{{\textstyle{d+1\over2}}} 
\def\ttd{{[\hadd]}}

\def\tp{{\tilde p}} 
\def\hap{{\textstyle{p\over2}}} 
\def\tpp{{[\hap]}}

\def\pl{\prod\limits}
 
\def\tD{{\hat D}}

\def\rank{{\rm rank\,}}
\font\verysmall=cmr5
\def\phr{\raise1.1pt\hbox{\verysmall x}\kern-8pt\supset}
\def\bhr{\raise1.1pt\hbox{\verysmall x}\kern-9pt\subset}

 \def\D{\Delta} \def\bbz{Z\!\!\!Z}

\def\bac{{C\kern-5.2pt I}} \def\bbc{{C\kern-6.5pt I}} 
\def\bbr{I\!\!R}  
\def\a{\alpha} \def\b{\beta} \def\g{\gamma} \def\d{\delta} 
\def\G{\Gamma} 
   
  \def\l{\lambda} 
    
\def\cc{{\cal C}}   \def\tmu{{\tilde \mu}} 
\def\hc{\hat{C}}  
\def\tc{\tilde{\cal C}} 
\def\tih{{C_\chi}}  \def\tch{{\tilde{\chi}}}

\def\htt{\hat{T}} \def\tt{{{\tau}}}

\def\cm{{\cal M}}  
 \def\cac{{\cal A}} 

    \def\cs{{\cal S}}  
\def\co{{\cal O}}  
\def\cg{{\cal G}}    
 \def\hd{{D}} 
  
  \def\s{\sigma} 
 \def\x{\xi}  
\def\lg{\langle} \def\rg{\rangle} 
 \def\y{\eta} 
 \def\llra{\longleftrightarrow}  
\def\lra{\leftrightarrow}  
\def\rra{\longrightarrow} \def\lla{\longleftarrow} 
\def\Lra{\Longrightarrow} 
\def\ra{\rightarrow} \def\vr{\vert}

\def\pd{\partial}

\def\tn{{\tilde n}} \def\tN{{\tilde N}} 
\def\t{\tau} 
\def\kc{K_\chi^\t}  \def\kci{K_\tch^\tt}  
    
\def\dia{~~$\diamondsuit$}  


\lref\Nai{M.A. Naimark, Trudy Moskov. Mat. Obshch. {\bf 8} 
(1959) 121; English translation: Am. Math. Soc. Transl. {\bf 36} 
(1964) 101.} 

\lref\Hir{T. Hirai, Proc. Jap. Acad. {\bf 42} (1965) 323.}

\lref\Kn{A.W. Knapp, {\it Representation Theory of Semisimple  
Groups (An Overview Based on Examples)}, 
(Princeton Univ. Press, 1986).} 

\lref\KnSt{A.W. Knapp, Ann. Math. {\bf 93} (1971) 489.}

\lref\HC{Harish-Chandra, ~Acta Math. {\bf 113} (1965) 241; 
ibid. {\bf 116} (1966) 1.} 

\lref\HS{W. Schmid,~  Rice University Studies, {\bf 56} 
(1970) 99;\nl R. Hotta,~  J. Math. Soc. Japan, {\bf 23} 
(1971) 384.}

\lref\DP{V.K. Dobrev and V.B. Petkova,~   
Rep. Math. Phys. {\bf 13} (1978) 233.} 

\lref\FlFr{M. Flato and C. Fronsdal, 
J. Math. Phys. {\bf 22} (1981) 1100.}

\lref\AFFS{E. Angelopoulos, M. Flato, C. Fronsdal and D. Sternheimer,
Phys. Rev. {\bf D23} (1981) 1278.}

\lref\Fro{ C. Fronsdal, Phys. Rev. {\bf D26} (1982) 1988.}

\lref\NiSe{H. Nicolai and E. Sezgin, Phys. Let. {\bf 143B} (1984) 103.}

\lref\TMP{I.T. Todorov, M.C. Mintchev and V.B. Petkova,~  
{\it Conformal Invariance in Quantum Field Theory}, 
(Pisa, Sc. Norm. Sup., 1978).} 

\lref\SoSt{G.M. Sotkov and D.Ts. Stoyanov, J. Phys. {\bf A13} 
(1980) 2807.} 

\lref\BFH{B. Binegar, C. Fronsdal and W. Heidenreich, 
J. Math. Phys. {\bf 24} (1983) 2828.} 

\lref\PeSo{V.B. Petkova and G.M. Sotkov, 
Lett. Math. Phys. {\bf 8} (1984) 217.} 

\lref\DPm{V.K. Dobrev and V.B. Petkova, 
Lett. Math. Phys. {\bf 9} (1985) 287.} 

\lref\DPf{V.K. Dobrev and V.B. Petkova, 
Fortschr. d. Phys. {\bf 35} (1987) 537-572.} 

\lref\DPu{V.K. Dobrev and V.B. Petkova, 
Phys. Lett. {\bf 162B} (1985) 127.}

\lref\DPp{V.K. Dobrev and V.B. Petkova, 
Proceedings, eds. A.O. Barut and H.D. Doebner, 
Lecture Notes in Physics, Vol. 261 (Springer-Verlag, Berlin, 1986) 
p. 291 
and p. 300.} 

\lref\GNW{M. Gunaydin, P. van Nieuwenhuizen and N.P. Warner, 
Nucl. Phys. {\bf B255} (1985) 63.} 

\lref\KRN{H.J. Kim, L.J. Romans and P. van Nieuwenhuizen,
Phys. Rev. {\bf D32} (1985) 389.}

\lref\GuMa{M. Gunaydin and N. Marcus, 
Class. Quant. Grav. {\bf 2} (1985) L11.}

\lref\Domu{V.K. Dobrev,~ Lett. Math. Phys. {\bf 9} 
(1985) 205.}

\lref\Malda{J. Maldacena,~  Adv. Theor. Math. Phys. {\bf 2} (1998) 231 
(hep-th/971120).}

\lref\FeFra{S. Ferrara and C. Fronsdal,~  Class. Quant. Grav. {\bf 15} 
(1998) 2153 (hep-th/9712239).}

\lref\GuMi{M. Gunaydin and D. Minic,~  
Nucl. Phys. {\bf B523} (1998) 145 
(hep-th/9802047).} 

\lref\GKP{S.S. Gubser, I.R. Klebanov and A.M. Polyakov,~  
Phys. Lett. {\bf 428B} (1998) 105 
(hep-th/9802109).} 

\lref\FeFrb{S. Ferrara and C. Fronsdal,
Phys. Lett. {\bf 433B} (1998) 19 
(hep-th/9802126).}

\lref\Wi{E. Witten,~  Adv. Theor. Math. Phys. {\bf 2} (1998) 253 
(hep-th/9802150).}

\lref\Lan{R.P. Langlands, On the classification of
irreducible representations of real algebraic groups,
Mimeographed notes, Princeton (1973).}

\lref\KnZu{A.W. Knapp and G.J. Zuckerman, in: Lecture Notes in Math. 
Vol. 587 (Springer, Berlin, 1977) pp. 138-159.; 
~Ann. Math. {\bf 116} (1982) 389-501.}

\lref\Mack{G. Mack,~  J. Funct. Anal. {\bf 23} (1976) 311.} 

\lref\DMPPT{V.K. Dobrev, G. Mack, V.B. Petkova, S.G. Petrova and I.T.
Todorov,~  {\it Harmonic Analysis on the ~ $n$ - Dimensional Lorentz}  
{\it Group and Its Applications to Conformal Quantum Field Theory}, 
Lecture Notes in Physics, Vol. 63 (Springer, 1977).} 

\lref\Dob{V.K. Dobrev,~ Rep. Math. Phys. {\bf 25} (1988) 159.} 

\lref\Dobp{V.K. Dobrev,~   in preparation.} 

\lref\Kol{K. Koller,~  Comm. Math. Phys. {\bf 40} (1975) 15.} 

\lref\FGGP{S. Ferrara, A.F. Grillo, R. Gatto and G. Parisi,~  
Lett. Nuovo Cim. {\bf 4} (1972) 115.} 

\lref\Mig{A.A. Migdal,~  4-dimensional soluble models of 
conformal field theory, preprint, Landau Inst. Theor. Phys., 
Chernogolovka (1972).} 

\lref\Pol{A.M. Polyakov,~  JETP Lett. {\bf 12} (1970) 381.} 

\lref\FrPa{E.S. Fradkin and M.Ya. Palchik, 
Nucl. Phys. {\bf B99} (1975) 317.}

\lref\RG{I.S. Gradshteyn and I.M. Ryzhik,~  
{\it Tables of Integrals, Series and Products} 
(Academic Press, NY, 1965).}

\lref\FMMRa{D.Z. Freedman, S.D. Mathur, A. Matusis and L. Rastelli,~  
MIT-CTP-2727, hep-th/9804058.}


\lref\HeSf{M. Henningson and K. Sfetsos, 
 Phys. Lett. {\bf 431B} (1998) 63   (hep-th/9803251).} 

\lref\MuVia{W. M\"uck and K.S. Viswanathan, Phys. Rev. {\bf D58} (1998) 
041901  (hep-th/9804035).}

\lref\LiTsa{H. Liu and A.A. Tseytlin,  Nucl. Phys. {\bf B533} (1998) 
88 (hep-th/9804083).} 

\lref\MeTs{R.R. Metsaev and A.A. Tseytlin, hep-th/9805028}

\lref\CNSS{G. Chalmers, H. Nastase, K. Schalm and R. Siebelink, 
   ITP-SB-98-36, hep-th/9805105.}

\lref\MuVib{W. M\"uck and K.S. Viswanathan, 
 Phys. Rev. {\bf D58} (1998) 106006 (hep-th/9805145).} 

\lref\CKPF{A.M. Ghezelbash, K. Kaviani, S. Parvizi and A.H. Fatollahi,  
 Phys. Lett. {\bf 435B} (1998) 291   (hep-th/9805162).}

\lref\LMRS{S. Lee, S. Minwalla, M. Rangamani and N. Seiberg, PUPT-1796,  
   hep-th/9806074.} 

\lref\Akh{E.T. Akhmedov, ITEP-TH-32-98, hep-th/9806217.}

\lref\CKA{A.M. Ghezelbash, M. Khorrami and A. Aghamohammadi,  
   hep-th/9807034.}

\lref\LiTsb{H. Liu and A.A. Tseytlin, IMPERIAL-TP-97-98-060, 
   hep-th/9807097.}

\lref\FMMRb{D.Z. Freedman, S.D. Mathur, A. Matusis and L. Rastelli, 
MIT-CTP-2770, hep-th/9808006.}

\lref\Corl{S. Corley, hep-th/9808184.}

\lref\Vol{A. Volovich, hep-th/9809009.}

\lref\Yi{W.S. l'Yi, HEP-CNU-9810, hep-th/9809132; 
HEP-CNU-9812, hep-th/9811097.}

\lref\HoFr{E. D'Hoker and D.Z. Freedman, hep-th/9809179.} 
     
\lref\BCFM{D. Berenstein, R. Corrado, W. Fischler and J. Maldacena, 
  UTTG-05-98, hep-th/9809188.} 

\lref\Bon{G. Bonelli,  SISSA-114-98-EP, hep-th/9810194.} 

\lref\OhZh{N. Ohta and J.-G. Zhou, hep-th/9811057.} 
 
\lref\BiKo{M. Bianchi and S. Kovacs, ROM2F-98-37, hep-th/9811060.} 
 
\lref\Liu{H. Liu, hep-th/9811152.}

\lref\DhFr{E. D'Hoker and D.Z. Freedman, UCLA/98/TEP/34, 
MIT-CTP-2795, hep-th/9811257.} 


\lref\FFZ{S. Ferrara, C. Fronsdal and A. Zaffaroni
Nucl. Phys. {\bf B532} (1998) 153  (hep-th/9802203).}

\lref\FlFrb{M. Flato and C. Fronsdal,  Lett. Math. Phys. 
{\bf 44} (1998) 249 (hep-th/9803013).} 

\lref\OzTe{Y. Oz and J. Terning,  Nucl. Phys. {\bf B532} 
(1998) 163   (hep-th/9803167).} 

\lref\AnFea{L. Andrianopoli and S. Ferrara,  Phys. Lett. {\bf 430} (1998) 
248  (hep-th/9803171).}

\lref\FLZ{S. Ferrara, M.A. Lledo and A. Zaffaroni, 
 Phys. Rev. {\bf D58} (1998) 105029 (hep-th/9805082).}

\lref\BKL{V. Balasubramanian, P. Kraus and A. Lawrence, 
   HUTP-98-A028,  (hep-th/9805171).}

\lref\GMZa{M. Gunaydin, D. Minic and M. Zagermann,  
Nucl. Phys. {\bf B534} (1998) 96 (hep-th/9806042).}

\lref\FeFrc{S. Ferrara and C. Fronsdal, CERN-TH-98-186, hep-th/9806072.}

\lref\FeZa{S. Ferrara and A. Zaffaroni, CERN-TH-98-229, 
   hep-th/9807090.}

\lref\AnFeb{L. Andrianopoli and S. Ferrara, CERN-TH-98-234, hep-th/9807150.}

\lref\OlRd{M.J. O'Loughlin and S. Randjbar-Daemi, 
ICTP preprint IC/98/100, hep-th/9807208.} 

\lref\FPZ{S. Ferrara, M. Porrati, A. Zaffaroni, CERN-TH-98-330, 
hep-th/9810063.}

\lref\GMZb{M. Gunaydin, D. Minic and M. Zagermann, PSU-TH-205, 
hep-th/9810226.}

\lref\AnFec{L. Andrianopoli and S. Ferrara, CERN-TH-98-388, 
hep-th/9812067.} 


\centerline{{\sfont Intertwining Operator Realization}}  
\vskip 3mm 
\centerline{{\sfont of the  AdS/CFT Correspondence}}

\vskip 1.1cm

\centerline{{\bf V.K. Dobrev}$^*$}
\footnote{}{$^{*}$ ~\fmale{~Permanent address: 
Bulgarian Academy of Sciences, 
Institute of Nuclear Research and Nuclear Energy, 
72 Tsarigradsko Chaussee, 1784 Sofia, Bulgaria.}}

\vskip 0.3cm

\centerline{Arnold Sommerfeld Institute for Mathematical Physics} 
\centerline{Technical University Clausthal} 
\centerline{Leibnizstr. 10, ~38678 Clausthal-Zellerfeld, ~Germany} 
\centerline{asvd@rz.tu-clausthal.de} 
\vskip 6mm

\centerline{{\bf Abstract}}

\midinsert\narrower\narrower{\male 
\nt We give a group-theoretic interpretation of 
the AdS/CFT correspondence as relation of ~representation 
equivalence~ between representations of 
the conformal group describing the bulk AdS fields 
~$\phi$, their boundary fields ~$\phi_0$~ and the 
coupled to the latter boundary conformal operators ~${\cal O}$.
We use two kinds of equivalences. The first kind is equivalence 
between the representations describing the bulk fields and the 
boundary fields and it is established  here. 
The second kind is the equivalence between conjugated conformal 
representations related by Weyl reflection, e.g., the coupled fields 
~$\phi_0$~ and ~${\cal O}$.  Operators realizing the first 
kind of equivalence for special cases  
were actually given by Witten and others - here they are 
constructed in a more general setting from the requirement that they are  
intertwining operators. 
The intertwining operators realizing the second kind of equivalence are  
provided by the standard conformal two-point functions.
Using both equivalences we find that the bulk field has in fact 
~two~ boundary fields, namely, the coupled fields 
~$\phi_0$~ and ~${\cal O}$, the limits being governed by the 
corresponding conjugated conformal weights ~$d-\Delta$ and ~$\Delta$. 
Thus,  from the viewpoint of the 
bulk-boundary correspondence the coupled fields $\phi_0$ and 
${\cal O}$ are generically on an equal footing. 
\hfil\break \indent
Our setting is more general since our bulk fields are described by 
representations of the  Euclidean conformal group, i.e., the 
de Sitter group ~$G=SO(d+1,1)$, 
induced from representations ~$\tau$~ of the maximal compact 
subgroup ~$SO(d+1)$~ of ~$G$. From these large reducible 
representations we can single out representations which 
are  equivalent to conformal boundary representations labelled 
by the conformal weight and by 
arbitrary representations ~$\mu$~ of the Euclidean Lorentz 
group ~$M=SO(d)$, such that ~$\mu$~ is contained in the restriction of 
~$\tau$~ to $M$.  Thus, our boundary~$\lra$~bulk operators 
can be compared with those in the literature only when for a fixed ~$\mu$~ we 
consider a 'minimal' representation  ~$\tau=\tau(\mu)$~ 
containing ~$\mu\,$. We also relate the boundary~$\ra$~bulk 
normalization constant to the Plancherel measure for $G$.
}\endinsert 


\np 
\baselineskip=12pt

\newsec{Introduction}

\nt 
Recently there was renewed interest in (super)conformal field 
theories in arbitrary dimensions. This happened after the 
remarkable proposal in \Malda, according to which 
the large $N$ limit of a conformally invariant theory in $d$ dimensions
is governed by supergravity (and string theory) on $d+1$-dimensional
 $AdS$ space  (often called $AdS_{d+1}$) times a compact manifold.
Actually the possible relation of field theory on $AdS_{d+1}$ 
to field theory on $\cm_d$ has been a subject of long interest, cf., 
e.g., \refs{\FlFr\AFFS\Fro\NiSe{--}\GNW}, and also  
\refs{\FeFra\GuMi{--}\FeFrb} for discussions motivated by 
recent developments, and additional references.
The proposal of \Malda$\,$ was elaborated in \GKP$\,$ and \Wi$\,$ where was 
proposed a precise correspondence between conformal field theory 
observables and those of supergravity:~ correlation functions in 
conformal field theory are given by the dependence of the supergravity 
action on the asymptotic behavior at infinity. 
More explicitly, a conformal field ~$\co$~ corresponds to an 
AdS field ~$\phi$~ when there exists a conformal invariant 
coupling ~$\int \phi_0\, \co$~ where ~$\phi_0$~ is the 
value of ~$\phi$~ at the boundary of $AdS_{d+1}\,$ \Wi.      
Furthermore, the dimension ~$\D$~ of the operator ~$\co$~ 
is given by the mass of the particle described by 
~$\phi$~ in supergravity \Wi.  Also the the spectrum
of Kaluza-Klein excitations of $AdS_5\times S^5$, as computed in
\refs{\KRN,\GuMa}, can be matched precisely
with certain operators of the $N=4$ super Yang-Mills 
theory in four dimensions \Wi. 
After these initial papers there was an explosion of 
related research of which of interest to us are two 
aspects: 1) calculation of conformal 
correlators from AdS (super)gravity, cf., e.g., \refs{\HeSf\MuVia\FMMRa
\LiTsa\CNSS\MuVib\CKPF\LMRS\CKA\LiTsb\FMMRb
\Corl\Vol\Yi\HoFr\BCFM\Bon\BiKo\Liu{--}\DhFr};\ 2) matching of 
gravity and string spectra with conformal theories, 
cf., e.g., \refs{\FFZ\FlFrb\OzTe\AnFea\FLZ\BKL\GMZa\FeFrc
\FeZa\AnFeb\OlRd\FPZ\GMZb{--}\AnFec}.

One of the main features furnishing this correspondence is that 
the boundary $\cm_d$ of $AdS_{d+1}$ is in fact
a copy of $d$-dimensional Minkowski space  (with a cone added 
at infinity); the symmetry group $SO(d,2)$ of $AdS_{d+1}$ acts
on $\cm_d$ as the conformal group.  The fact that 
$SO(d,2)$ acts on $AdS_{d+1}$ as a group of ordinary symmetries 
and on $\cm_d$ as a group
of conformal symmetries means that  there are two ways to
get a physical theory with $SO(d,2)$ symmetry:
in a relativistic field theory (with or without gravity) on 
$AdS_{d+1}$, or in a conformal field theory on $\cm_d$.  

The main aim of this paper is to give a group-theoretic interpretation 
of the above correspondence. In fact such an interpretation is partially 
present in \DMPPT$\,$ for the $d=3$ Euclidean version of the 
AdS/CFT correspondence in the context of the construction of discrete 
series representations of the  group $SO(4,1)$ involving 
symmetric traceless tensors of arbitrary nonzero spin. 

In short the essence of our interpretation is that the above 
correspondence is a relation of ~{\it representation 
equivalence}~ between the representations describing the fields 
~$\phi$, ~$\phi_0$~ and ~$\co$. There are actually two kinds of 
equivalences. The first kind is new (besides the example from 
\DMPPT$\,$ mentioned above) and is proved here - it is 
between the representations describing the bulk fields and the 
boundary fields. The second kind  is well known - it is the equivalence 
between boundary conformal representations  which are 
related by Weyl reflections, the representations here being the 
coupled  fields ~$\phi_0$~ and  ~$\co$. 
  
Our interpretation means that the operators relating these fields 
are ~{\it intertwining operators}~ between 
(partially) equivalent representations.  Operators giving the first 
kind of equivalence for special cases were actually given in 
\refs{\Wi,\HeSf,\LiTsa\CNSS{--}\MuVib,\Corl\Vol{--}\Yi} 
- here they are constructed in a more general setting 
from the requirement that they are  intertwining 
operators. The operators giving the second kind of equivalence are  
provided by the standard conformal two-point 
functions (and the latter intertwining property was known 
long time ago, cf. \Kol,  also  \DMPPT). 
Using both equivalences we find that the bulk field has naturally 
~{\it two}~ boundary fields, namely, the coupled fields 
~$\phi_0$~ and ~${\cal O}$, the limits being governed by the 
corresponding conjugated conformal weights ~$d-\D$ and ~$\D$. 
Thus, we notice that from the point of view of the 
bulk-to-boundary correspondence 
the coupled fields ~$\phi_0$~ and ~${\cal O}$ are 
generically\ {\it on an equal footing.}  

As mentioned, {\it our setting is  more general}.  
In order to be more specific we consider here the 
Euclidean version of the AdS/CFT correspondence. In this 
case the conformal group is the ~{\it de Sitter 
group} ~$G=SO(d+1,1)$. 
Our bulk fields are obtained from representations of ~$G$~ 
induced from representations ~$\tau$~ of the maximal compact 
subgroup ~$K=SO(d+1)$~ of ~$G$. From these large reducible 
representations we can single out representations which 
are  equivalent to conformal boundary representations labelled 
by arbitrary representations ~$\mu$~ of the Euclidean Lorentz 
group ~$M=SO(d)$, such that ~$\mu$~ is contained in the restriction of 
~$\t$~ to $M$. Thus, our boundary~$\lra$~bulk operators 
can be compared with those in the literature only when for a fixed ~$\mu$~ we 
consider a 'minimal' representation  ~$\tau=\tau(\mu)$~ 
containing ~$\mu\,$. We also relate the boundary~$\ra$~bulk 
normalization constant to the Plancherel measure for $G$.

On the AdS side the representations are realized on 
~{\it de Sitter space}~$\cs$ (the Euclidean counterpart of AdS space), 
which is isomorphic to the coset $G/K$. 
What is also very essential for our approach is that ~$\cs$~ is 
isomorphic (via the Iwasawa decomposition) also 
to the solvable product group  
~$\tN A$, where ~$\tN$~ is the abelian group of 
Euclidean translations  (isomorphic to ~$\bbr^d$), 
~$A$~ is the one-dimensional dilatation group.  
The isomorphism ~$\cs\cong\tN\,A$~ and related ones 
are explicated in Section 2. 
On the conformal side the representations are 
realized on ~$\tN$, and we use also the fact that the latter is locally 
isomorphic  (via the Bruhat decomposition) 
to the coset ~$G/MAN$, (where ~$N$~ is the group of 
special conformal transformations). These representations, 
called elementary representations (ERs), are introduced 
in Section 3. The representations on de Sitter space are
introduced in Section 4. Then in Section 5 we give the 
bulk-to-boundary and boundary-to-bulk intertwining 
operators and discuss the difference between equivalence 
and partial equivalence. There we display the second limit  
of the bulk fields and we derive some further intertwining 
relations. From the latter we derive the relation 
to the Plancherel measure for $G$. We illustrate the intertwining 
relations by a commutative diagram. 
Sections 6 contains some more comments and outlook. 

\np 
\vskip 5mm 

\newsec{de Sitter space from Iwasawa decomposition} 

\nt 
As we mentioned the relation between the two pictures uses the fact that 
$\bbr^d$ is easily identified within the $d+1$-dimensional 
de Sitter space. Indeed, de Sitter space may be parametrized as:
\eqn\dss{ \x_{d+2}^2 - \sum_{\a=1}^{d+1} \x_\a^2 ~=~ 1 ~, 
\quad \x_{d+2} \geq 1} 
and the first ~$d$~ of the ~$\x_\a$~ may be taken as 
coordinates on $\bbr^d$. 

The group-theoretic interpretation of this relation is present in \DMPPT$\,$ 
using the so-called ~{\it Iwasawa}~ decomposition 
~$G ~=~ \tN AK$.\foot{There are several versions of the 
Iwasawa decomposition, e.g., one may use the group $N$ instead of 
$\tN$, and one may take different order of the three factors involved. 
The choice of version is a matter of convenience.} 
 This is a global decomposition, i.e., each 
element $g$ of $G$ can be decomposed as the product of the corresponding 
matrices: ~$g = \tn a k$, with $\tn \in \tN$, $a\in A$, $k\in K$.  
To be explicit we use the following defining relation of $G$~: 
\eqn\dsg{ G ~=~ \{ g \in GL(d+2, \bbr) ~\vert ~ 
^t g\y g = \y \doteq {\rm diag}(-1,\dots,-1,1), 
~~\det g =1, ~~g_{d+2,d+2} \geq 1 \} }
which is the identity component of ~$O(d+1,1)$, 
($^t g$ is the transposed of $g$). Then we have 
the following matrix representations of the 
necessary subgroup elements (cf. formulae (1.20a), (1.21), (1.23), 
of \DMPPT, with $2h$ replaced here by $d$)~:
\eqna\mtra
$$\eqalignno{
\tn ~=~ \tn_x ~=&~ \pmatrix{ \d_{i j} &-x_i & x_i \cr  
x_j & 1-\half x^2 & \half x^2 \cr 
x_j & -\half x^2 &1+ \half x^2 \cr} \in\tN ~, \quad x\in \bbr^d,  
\quad x^2 \equiv \sum_{i=1}^d x_i^2 
& \mtra a\cr
a ~=&~ \pmatrix{ \d_{i j} & 0 & 0 \cr  
0 & { \vr a\vr^2 +1 \over 2 \vr a\vr } & 
{ \vr a\vr^2 -1 \over 2 \vr a\vr } \cr 
0 & { \vr a\vr^2 -1 \over 2 \vr a\vr } & 
{ \vr a\vr^2 +1 \over 2 \vr a\vr } } 
 \in A \, , \quad \vr a \vr >0 
& \mtra b\cr
k ~=&~ \pmatrix{ k_{i j} & k_{i,d+1} & 0 \cr 
k_{d+1,j} & k_{d+1,d+1} & 0 \cr 
0&0&1 \cr} \in K ~, \quad \left( k_{\a\b} \right) \in SO(d+1) 
& \mtra c\cr
}$$ 

Writing ~$g = \tn a k$~ one may determine the factors ~$\tn, a, k$~   
in terms of the matrix elements of $g$.  We use this for the 
elements of the last column of $g$, which actually parametrize 
the de Sitter space, i.e., 
\eqn\dsa{\x_A ~=~ g_{A,d+2} ~, \quad A = 1,\dots,d+2 ~.} 
Indeed, substituting \dsa$\,$ in \dss$\,$ we recover the 
~$(d+2,d+2)$-element of the defining relation \dsg, i.e., 
\eqn\dsgg{\eqalign{
&(^t g\y g)_{d+2,d+2}  ~=~ \y_{d+2,d+2} ~\Lra \cr  
& g_{d+2,d+2}^2 - \sum_{a=1}^{d+1} g_{a,d+2}^2 ~=~ 1 }} 

Now in terms of the parameters in \mtra{} we get for the 
elements of the last column of $g$, resp., for the parameters of 
de Sitter space~: 
\eqn\sll{ \eqalign{ 
g_{i,d+2} ~=&~ \xi_i ~=~ {1\over \va}\, x_i  \cr 
g_{d+1,d+2} ~=&~ \xi_{d+1} ~=~ { \vr a\vr^2 -1 + x^2\over 2 \vr a\vr } 
\cr 
g_{d+2,d+2} ~=&~  \xi_{d+2} ~=~ { \vr a\vr^2 +1 + x^2\over 2 \vr a\vr }
~\geq 1 }}
Notice that the only the ~$d+1$~ parameters of the matrices 
~$\tn_x\,,a$~ of the Iwasawa decomposition (cf. \mtra{a,b}) 
are involved.  Solving \sll$\,$ we obtain for the latter parameters: 
\eqn\sol{ \eqalign{
x_i ~=&~ {g_{i,d+2} \over g_{d+2,d+2} - g_{d+1,d+2}}
 ~=~ {\x_i \over \x_{d+2} -\x_{d+1}} \cr 
\vr a \vr ~=&~ {1\over g_{d+2,d+2} - g_{d+1,d+2}} ~=~ 
{1\over \x_{d+2} - \x_{d+1}} ~>~ 0
}}
(The last condition follows from $g_{d+2,d+2}\geq 1$.) 
From \sll$\,$ we get also the consistency condition:
\eqn\cst{ {\va^2 + x^2 \over \vr a \vr} ~=~ g_{d+2,d+2} + g_{d+1,d+2}
 ~=~ \x_{d+2} + \x_{d+1}} 
Indeed, inserting \sol$\,$ in \cst$\,$ we recover \dss$\,$ and \dsgg. 

\medskip

\item{\bu} Thus, in \sll$\,$ and \sol$\,$ we have the mentioned correspondence 
between de Sitter space ~$\cs$~ and the (coset) solvable subgroup 
~$S ~=~ \tN A ~\cong~ G/K ~\cong \bbr^d \times \bbr_{>0}$~ 
of the de Sitter group $G$. In addition, this explicates 
the group-theoretical interpretation of Euclidean space-time ~$\bbr^d$~ 
as the abelian subgroup ~$\tN$~ of the solvable subgroup ~$S$, and the 
topological interpretation of ~$\bbr^d$~ as the boundary of 
~$\bbr^d \times \bbr_{>0}$~ for ~$\va\to 0$. 

\medskip

\nt{\it Remark:}~~ Note that for ~$d~~even$~ some expressions 
are simpler if we work with the extended de Sitter group:
\eqn\ddsg{ G' ~\doteq~ \{ g \in GL(d+2, \bbr) ~\vert ~ 
^t g\y g = \y, ~~g_{d+2,d+2} \geq 1 \} }
which includes reflections of the first $d+1$ axes. 
Then ~$K\to K'=O(d+1)$, ~$M\to M'=O(d)$, but the de Sitter space 
~$\cs$~ and the results of this Section are not changed.\dia 

\np 

\vskip 5mm \newsec{Conformal field theory representations}

\nt 
The representations  used in conformal field theory 
are called (in the representation theory of semisimple Lie groups) 
generalized principal series representations  (cf. \Kn). 
In \DMPPT, \DP, \Dob$\,$  
they were called ~{\it elementary representations} (ERs). 
They are obtained by induction from the subgroup$\,$ $P=MAN$, 
($P$ is called a parabolic subgroup of $G$). The induction is from 
unitary irreps of $M=SO(d)$, from arbitrary (non-unitary) characters of $A$, 
and trivially from $N$. There are several realizations of these 
representations. We give now the so-called ~{\it noncompact picture}~ 
of the ERs - it is the one actually used in physics. 

The representation space of these induced representations 
consists of smooth functions on ~$\bbr^d$~ with values in 
the corresponding finite-dimensional representation space of ~$M$, i.e.: 
\eqn\funn{ C_\chi ~=~ \{ f \in C^\infty(\bbr^d,V_\mu)\} }  
where$\,$ $\chi\, =\, [\mu,\D]$,$\,$ ~$\D$~ is the conformal weight, 
~$\mu\,$ is a unitary irrep of $M$,$\,$   
$V_\mu\,$ is the finite-dimensional representation space of $\mu$. 
In addition, these functions have special 
asymptotic expansion as ~$x\to \infty$. The leading term 
of this expansion is ~$f(x) ~\sim~ {1\over (x^2)^\D}\, f_0$, 
(for more details of this expansion we refer to \DMPPT, \DP, \Dob). 
The representation ~$T^\chi$~ acts in $\tih$ by: 
\eqn\lart{(T^\chi(g) f) (x) ~=~ \va^{-\D} \cdot  \hd^\mu(m)\,  
f ( x') } 
where the nonglobal Bruhat decomposition $g=\tn m a n$ is 
used:\foot{For the cases with measure zero for which ~$ g^{-1}\tn_x$~ 
does not have a Bruhat decomposition of the form $\tn man$ 
the action is defined 
separately, and the passage from \lart$\,$ to these special 
cases is ensured to be smooth by the asymptotic 
properties mentioned above. Further, we may omit such  
measure-zero exceptions from the formulae - in a rigorous 
exposition all of them are taken care of, cf. \DMPPT, \DP, \Dob.} 
\eqn\nnam{  g^{-1}\tn_x ~=~ \tn_{x'} m^{-1}{a}^{-1} n^{-1}  ~, \quad 
g\in G , \, \tn_x, \tn_{x'} \in \tN , \, m\in M,\, a\in A,\, n\in N}
$\hd^\mu(m)$ is the representation matrix of $\mu$ in $V_\mu\,$.\foot{
One may interpret  ~$C_\chi$~ also 
as a space of smooth sections of the homogeneous
vector bundle with base space ~$G/MAN$~ and fibre ~$V_\mu\,$.}

The ERs are generically irreducible both operatorially (in the 
sense of Schur's Lemma) and topologically (meaning nonexistence 
of nontrivial (closed) invariant subspaces). However, 
most of the physically relevant examples are when the ERs are 
topologically reducible and indecomposable.
In particular, such are the representations describing gauge 
fields, cf., e.g., \refs{\TMP\SoSt\BFH{--}\PeSo}.

The importance of the elementary representations comes also from the 
remarkable result of Langlands-Knapp-Zuckerman 
\Lan, \KnZu$\,$ stating that every irreducible admissible representation of a
real connected semisimple Lie group $G$ with finite centre is
equivalent to a subrepresentation of an elementary representation 
of $G$.\foot{{\it Subrepresentations}~ 
are irreducible representations realized on 
invariant subspaces  of the ER spaces  
(in particular, the irreducible ERs themselves).
The admissibility condition is fulfilled in the 
physically interesting examples.} 
To obtain a subrepresentation of a topologically reducible ER 
one has to solve certain invariant differential equations, 
cf. \DMPPT, \DP, \Dob. 

Note that the representation data given by ~$\chi=[\mu,\D]$~ fixes also 
the value of the Casimir operators ~$\cc_i$~ 
in the  ER$\,$ $C_\chi\,$, independently of the latter reducibility. 
For later use we write:
\eqn\cas{\cc_i(\{ X\})\, f(x) ~=~ \l_i(\mu,\D)\, f(x) ~, \qquad 
i=1,\dots,\rank G ~=~ [\had]+1,}
where ~$\{X\}$~ denotes symbolically the generators 
of the Lie algebra ~$\cg$~ of ~$G$, and the action of ~$X\in\cg$~ 
is given by the infinitesimal version of \lart:
\eqn\larf{ (X\, f) (x) ~\doteq~ {\pd \over \pd t}\, 
(T^\chi(\exp t X) f) (x)\vert_{t=0}}  
applying the Bruhat decomposition to ~$\exp (-t X)\, \tn_x\,$.

Next, we would like to recall the general expression of 
the conformal two-point function ~$G_\chi(x_1-x_2)$~ 
(for special cases cf. \refs{\Pol\FGGP\Mig{--}\FrPa}, for the 
general formula with special stress on the role of the 
conformal inversion, cf. \Kol, also \DMPPT):
\eqn\tpf{ \eqalign{ 
G_\chi(x) ~=&~ {\g_\chi\over (x^2)^\D }\, \hd^\mu(r(x)) \cr &\cr 
r(x) ~=&~ \pmatrix{\tilde r(x) 
& 0 & 0 \cr    
0 & 1 & 0 \cr  0 & 0 & 1 } \in M ~, \quad 
\tilde r(x) ~=~ \left( {2\over x^2}\, x_i x_j - \d_{i j} \right) 
}}
where ~$\g_\chi$~ is an arbitrary constant for the moment. 
(Note that for ~$d~~even$~ $r(x)\in O(d)$, so we work with ~$G'$, 
cf. \ddsg.)  

Finally, we note the 
intertwining property of ~$G_\chi(x)$. Namely, let ~$\tch$~ be 
the representation conjugated to ~$\chi$~ by 
Weyl reflection, i.e., by the nontrivial element of the two-element 
restricted Weyl group $W(G,A)$ \DMPPT. Then we have: 
\eqn\chcj{ \tch ~\doteq~ [\, \tmu,\, d-\D\, ] ~, \quad {\rm for} ~
\chi = [\mu,\D], } 
where ~$\tmu$~ is the ~{\it mirror image}~ representation of 
~$\mu$. (For ~$d$~ odd ~$\tmu \cong \mu$, while for ~$d$~ even ~$\tmu$~ 
may be obtained from ~$\mu$~ by exchanging the representation 
labels of the two distinguished Dynkin nodes of ~$SO(d)$.)   
Then there is the following intertwining operator \Kol, \DMPPT:
\eqna\inti
$$\eqalignno{&G_\chi ~:~ C_\tch \rra C_\chi ~, \qquad 
T^\chi (g) \circ G_\chi ~=~ G_\chi \circ T^\tch (g) ~, \quad \forall g ~, 
&\inti a\cr 
&(G_\chi f) (x_1) ~\doteq ~ \int G_\chi(x_1-x_2)\, f(x_2)\, d x_2  ~.
&\inti b\cr }$$ 
($d x \equiv d^d x$) 
which means that the representations are partially equivalent. 
Note that because of this equivalence the 
values of all Casimirs coincide:
\eqn\csmr{ \l_i(\tmu,  d-\D) ~=~ \l_i(\mu,  \D) ~, \qquad \forall i ~.} 

Note, that at generic points the representations are equivalent, namely, 
one has \DMPPT, \DP:  
\eqn\nrm{ G_\chi\, G_\tch ~=~ \id_\chi ~, \qquad 
G_\tch\, G_\chi\, ~=~ \id_\tch } 
This may be used to fix the constant ~$\g_\chi\,$. 

From the point of view of the AdS/CFT correspondence 
the importance of the pair ~$\chi,\tch$~ is in the 
fact that the corresponding fields have conformally 
invariant coupling through the standard bilinear form:
\eqn\frm{ \lg \phi_0\,, \co\rg ~\doteq~ \int d x\, 
\lg \phi_0(x)\,, \co(x)\rg_\mu ~, \qquad  \phi_0 \in C_\tch ~, 
~~~\co \in C_\chi ~, }
where ~$\lg \cdot,\cdot\rg_\mu$~ is the standard pairing between 
~$\mu$ and $\tmu$. (Note that if $\phi_0$ determines a $p$-form, 
then $\co(x)$ determines a $(d-p)$-form.)


\vskip 5mm \newsec{Representations on de Sitter space} 

\nt 
In the previous section we discussed representations 
on ~$\bbr^d\cong \tN$~ induced from the parabolic subgroup ~$MAN$~ 
which is natural since the abelian subgroup ~$\tN$~ is locally 
isomorphic to the factor 
space ~$G/MAN$ (via the Bruhat decomposition). 
Similarly, it is natural to discuss 
representations on de Sitter space ~$\cs\cong \tN A$~ 
which are induced from the maximal compact subgroup ~$K=SO(d+1)$~ 
since the solvable group ~$\tN A$~ is  isomorphic to the factor space ~$G/K$ 
(via the Iwasawa decomposition). Namely, we consider 
the representation space:
\eqn\funk{ \hc_\t   ~=~ \{ \phi \in 
C^\infty(\bbr^d\times\bbr_{>0}\,,U_\t) \} } 
where$\,$ $\t\,$ is an arbitrary unitary irrep of $K$,$\,$    
$U_\t\,$ is the finite-dimensional representation space of $\t$, 
with representation action:
\eqn\lartu{
(\htt^\t(g)\phi) (x,\vert a\vert) ~=~ \tD^\t (k)\, \phi ( {x'}, 
\vert a'\vert) } 
where the Iwasawa decomposition is used:  
\eqn\nak{  g^{-1}\tn_x a ~=~ \tn_{x'} a' k^{-1}  ~, \quad 
g\in G ,\, k\in K, \, \tn_x, \tn_{x'} \in \tN , \, a, a' \in A}
and $\tD^\t(k)$ is the representation matrix of $\t$ in $U_\t\,$. 
However, unlike the ERs, these representations are reducible, 
and to single out an  irrep equivalent, say, a 
subrepresentation of an ER, one has to look for solutions 
of the eigenvalue problem related to the Casimir operators. This 
procedure is actually well understood and used in the 
construction of the discrete series of unitary representations, 
cf. \HC, \HS, (also \DMPPT$\,$ for $d=3$). 

In the actual implementation of \lartu$\,$ it is convenient 
to use the unique decomposition:
\eqn\dkk{
k ~=~ m(k) k_f ~, \qquad 
m(k)=\pmatrix{\tilde m(k)& 0 &0 \cr 0&1&0 \cr 0&0&1} \in M ~, \qquad 
k_f=\pmatrix{\tilde k_f& 0  \cr 0&1} \in K }
representing the decomposition of 
~$K$~ into its subgroup ~$M$~ and the coset ~$K/M$~: ~
~$K ~\cong ~ M~K/M$. Explicitly, 
using the parametrization of ~$k$~ in \mtra{c}, 
we have (for $k_{d+1,d+1}\neq -1$):
\eqn\dek{\eqalign{ 
\tilde m(k) ~=&~ \left( k_{i j} - 
{1\over 1+k_{d+1,d+1}}\, k_{i,d+1}\, k_{d+1,j}  
\right) \cr 
\tilde k_f ~=&~ 
\pmatrix{\d_{i j} - {1\over 1+k_{d+1,d+1}}
k_{d+1,i} k_{d+1,j} & -k_{d+1,i}\cr 
k_{d+1,j} & k_{d+1,d+1}} ~=\cr 
=& \pmatrix{ \d_{i j}  - {2\over 1+x^2}\, x_{i} x_{j} 
& - {2\over 1+x^2}\, x_{i}\cr 
{2\over 1+x^2}\, x_{j} & {1-x^2\over 1+x^2}} ~\doteq ~\tilde k_x \cr 
x\in & ~\bbr^d ~, \quad 
x_i ~=~ {1\over 1+k_{d+1,d+1}}\, k_{d+1,i} ~, \quad 
x^2 ~=~ {1-k_{d+1,d+1} \over 1+k_{d+1,d+1} }}} 
Note that ~$k_x ~\doteq ~\pmatrix{\tilde k_x& 0  \cr 0&1}$~ 
appeared in (1.30a) of \DMPPT.\foot{
The matrices ~$k_x$~ realize the (local) isomorphisms:~
$\bbr^d~\cong ~\tN$ ~${ {\rm loc}\atop {\cong\atop \phantom{\rm loc} }} ~
K/M ~\cong~ G/MAN$ (using for the last isomorphism 
the Iwasawa decomposition in the version ~$G=KAN$).}  

Further, we would like to extract from ~$\hc_\t$~ 
a representation that may be equivalent to ~$C_\chi\,$, $\chi=[\mu,\D]$. 
The first condition for this is that the ~$M$-representation  ~$\mu$~ is 
contained in the restriction of the ~$K$-representation ~$\t$~ to ~$M$. 
Another condition  is that the two representations 
would have the same Casimir values ~$\l_i(\mu,\D)$. Having in mind 
the degeneracy of Casimir values for partially equivalent 
representations (e.g., \csmr) we add also the appropriate 
asymptotic condition. Furthermore, from now on we shall suppose 
that ~$\D$~ is real. Thus, we shall use the representations: 
\eqn\funr{\eqalign{ 
 \hc^\t_\chi ~=~ \{\, \phi \in \hc_\t ~:&~~ 
\cc_i(\{\hat X\})\, \phi (x,\va) ~=~ \l_i(\mu,\D)\, \phi (x,\va) ~,
\quad \forall i ~, \quad 
\mu\in\t\vert_M ~, \cr 
&\phi (x,\va) ~\sim~ \va^\D\, \varphi(x) ~ {\rm for}~ \va\to 0\,   
\} }}
where ~$\{\hat X\}$~ denotes symbolically the generators 
of the Lie algebra ~$\cg$~ with the action ~$\hat X$~ of ~$X\in\cg$~ 
given by the infinitesimal version of \lartu:
\eqn\larf{ (\hat X\, \phi) (x,\va) ~\doteq~ {\pd \over \pd t}\, 
(\htt^\t (\exp t X) \phi) (x,\va)\vert_{t=0}}  
applying the Iwasawa decomposition to ~$\exp (-t X)\,\tn_x\,a\,$. 
Certainly, the Casimirs ~$\cc_i(\{\hat X\})$~ with \larf$\,$ substituted 
are  differential operators and the elements of 
~$\hc_\chi$~ are  solutions of the equations above. 

\nt{\it Remark:}~~ Note that generically  
the functions in \funr$\,$ have also 
a second limit with ~$\D\to d-\D$~: 
\eqn\bndd{\tilde \varphi(x) ~=~ \lim_{\va\to 0}\, \va^{\D-d}\,  
\phi (x, \va) }
which will appear as a consequence of the formalism. 
With this we shall establish - for generic representations -  
the following important relation:
\eqn\dud{ \hc^\t_\chi ~=~ \hc^\t_\tch ~, \qquad \chi =[\mu,\D], ~
\tch =[\tmu,d-\D] ~, } 
for which besides \bndd, we use the equality between 
the Casimirs \csmr, and the fact that if ~$\t$~ contains ~$\mu$~ 
then it also contains the mirror image ~$\tmu\,$. 
This is established towards the end of 
next Section, where also some comment on the 
exceptional cases is made.\dia 

We end this Section by noting that for the representations on 
de Sitter space ~$\hc_\chi\,$,  ~$\hc_\t\,$, 
there is no exhaustivity result as the Langlands-Knapp-Zuckerman 
result \Lan, \KnZu$\,$ for ERs cited above. Thus, it is not surprising 
that not all conformal representations can be realized on de Sitter 
space, or, in other words, that some conformal fields live only 
on the boundary of de Sitter space and can not propagate into 
the bulk.


\vskip 5mm \newsec{Intertwining relations between conformal and 
de Sitter representations}

\newsubsec{Bulk-to-boundary intertwining relation} 

\nt
This Section contains our main results, explicating the relations between 
CFT and de Sitter representations as intertwining relations. 
We first give in this subsection the intertwining operator 
from the de Sitter to the CFT realization. 
The operator which we use is mapping a 
function on de Sitter space to its boundary value and was 
used in a restricted sense (explained below)  
in many papers, starting from \Wi. Also for those cases our 
result is new since we use it as operator between exactly 
defined spaces, and most importantly that we give it the 
interpretation of an intertwining operator. 

\nt 
{\bf Theorem:}~~ {\it Let us define the operator: 
\eqn\aaa{ L_\chi^\t ~:~  \hc^\t_\chi ~\rra ~C_\chi \, , }
with the following action:
\eqn\aab{ (L_\chi^\t \phi ) (x) ~=~ \lim_{\va\to 0}\ 
\va^{-\D}\ \Pi^\t_\mu\ \phi (x, \va) }
where ~$\Pi^\t_\mu$~ is the standard projection operator from the 
~$K$-representation space ~$U_\t$~ to the ~$M$-representation space
~$V_\mu\,$, which acts in the following way on the 
$K$-representation matrices: 
\eqn\prj{\Pi^\t_\mu\ \tD^\t (k) ~=~ \hd^\mu (m(k))\   \Pi^\t_\mu \ 
\tD^\t (k_f) } 
where we have used \dkk. 
Then ~$L_\chi^\t$~ is an intertwining operator, i.e.: 
\eqn\aac{L_\chi^\t   \circ  \htt^\t (g) ~=~ 
 T^\chi (g) \circ L_\chi^\t ~, \quad \forall g\in G ~.} 
In addition, in \aab\  the operator ~$\Pi^\t_\mu$~ 
acts in the following truncated way:}  
\eqn\prja{\Pi^\t_\mu\ \tD^\t (k) ~=~ \hd^\mu (m(k))\   \Pi^\t_\mu }
{\bf Proof:}~~ Applying the LHS side of \aac$\,$ to ~$\phi$~ we have:  
\eqna\prf
$$\eqalignno{ 
(L_\chi^\t   \circ  \htt^\mu (g) \phi ) (x) ~&=~ \lim_{\va\to 0}\
\va^{{-\D}}\ \Pi^\t_\mu \ ( \htt^\mu (g) \phi) (x, \va) ~=\cr &=~  
\lim_{\va\to 0}\ \va^{{-\D}}\ \hd^\mu (m(k))\ \Pi^\t_\mu\ \tD^\t(k_f)\ 
\phi (x', \vert a'\vert) 
~, &\prf {a} \cr &g^{-1}\tn_x a ~=~ \tn_{x'} a' k^{-1} &\prf {b}
}$$
Applying the RHS side of \aac$\,$ to ~$\phi$~ we have:
\eqna\prff
$$\eqalignno{ 
( T^\chi  (g)\circ L_\chi^\t \phi ) (x) ~&=~ 
\vr a''\vr^{-\D}\   \hd^\mu(m)\ (L_\chi^\t \phi) ( x'') ~=\cr 
&=~ \vert a''\vert^{-\D} \   \hd^\mu(m)\ 
\lim_{\va\to 0}\ \va^{{-\D}}\ \Pi^\t_\mu\ \phi (x'', \va) 
~,  &\prff {a} \cr 
&g^{-1}\tn_x ~=~ \tn_{x''} m^{-1}{a''}^{-1} n^{-1} 
&\prff {b}  }$$
In view of the Bruhat decomposition it is enough to prove
coincidence between \prf{a} and \prff{a} for ~$g ~=~ \tn_y\in\tN,
\ha\in A, \hm\in M$, and some element ~$w\in K$, $w\notin M$, ~representing 
the nontrivial element of the restricted Weyl group ~$W(G,A)$, 
since this element transforms elements of ~$\tN$~ into elements of
~$N$~: ~$w\tn_y w ~=~ n_{y'}$ (and thus makes unnecessary the 
check  ~$g=n_y$). Such an element is, e.g., $w =$diag$(1,\dots,1,-1,-1,1)$, 
i.e., rotation in the plane ~$(d,d+1)$. 
However, for simplicity we shall demonstrate only the case 
of odd ~$d$~ when we can take ~$w\to R$~ since in this case 
the conformal inversion ~$R ~\doteq~ $diag$(-1,\dots,-1,1)$~ is 
an element of ~$K$, (or we should suppose that we work 
with $G'$). We have: 

\item{\bu} ~$g=\tn_y$~: ~~then  \prf{b} gives ~$\tn_y^{-1}\tn_x a ~=~ 
\tn_{x-y} a$, i.e., ~$x'=x-y$, ~$a'=a$, ~$k=\id$, and 
(noting $\tD^\t(\id) = \id_\t$)~  \prf{a} becomes: 
\eqn\chk{\lim_{\va\to 0}\ \va^{{-\D}}\ \Pi^\t_\mu\ \phi (x-y, \va) }
while \prff{b} gives ~$\tn_y^{-1}\tn_x  ~=~ 
\tn_{x-y} $, i.e., ~$x''=x-y$, ~$m=\id$, ~$a''=\id$, ~$n=\id$, 
and (noting $\hd^\mu(\id) = \id_\mu$) 
\prff{a} also becomes \chk. 

\item{\bu} ~$g=\ha$~: ~~then  \prf{b} gives ~$\ha^{-1}\tn_x a ~=~ 
\tn_{{x\over\vert\ha\vert}}\ha^{-1} a$, ~i.e., ~$x'={x\over\vert\ha\vert}$, 
~$a'=\ha^{-1}a$, ~$k=\id$, and \prf{a} becomes:
\eqn\chka{\lim_{\va\to 0}\ \va^{{-\D}}\ \Pi^\t_\mu\ \phi
({x\over\vert\ha\vert},\ {\va\over\vert\ha\vert}) ~=~ \vert\ha\vert^{{-\D}}\
\lim_{\va\to 0}\ \va^{{-\D}}\ \Pi^\t_\mu\ \phi 
({x\over\vert\ha\vert},\ \va) } 
while \prff{b} gives ~$\ha^{-1}\tn_x  ~=~ 
\tn_{{x\over\vert\ha\vert}}\ha^{-1}$, ~i.e., ~$x''={x\over\vert\ha\vert}$, 
$m=\id$, ~$a''=\ha$, ~$n=\id$, and \prff{a} also becomes \chka.

\item{\bu} ~$g=\hm$~: ~~then  \prf{b} gives ~$\hm^{-1}\tn_x a ~=~ 
\tn_{\hm^{-1}x}a\hm^{-1}$, ~i.e., ~$x'_i = m^{-1}_{i j}x_j$, 
~$a'=a$, ~$k=m(k)=\hm$, and \prf{a} becomes:
\eqn\chkb{ \lim_{\va\to 0}\ \va^{{-\D}}\  
\Pi^\t_\mu\ \tD^\t(\hm)\  \phi(x', \va) ~=~ 
\hd^\mu(\hm)\  \lim_{\va\to 0}\ \va^{{-\D}}\  
\Pi^\t_\mu\ \phi(x', \va) } 
while \prff{b} gives ~$\hm^{-1}\tn_x  ~=~ 
\tn_{\hm^{-1}x}\hm^{-1}$, ~i.e., ~$x''= x'$, ~$m=\hm$, 
~$a''=\id$, ~$n=\id$, 
and \prff{a} also becomes \chkb.

\item{\bu} ~$g=R$~: ~~then  \prf{b} gives ~$R\tn_x a ~=~ 
\tn_{x'} a' k^{-1}$ with:
\eqn\exs{\eqalign{ 
&x' ~=~ x(x,a) ~\equiv~ - {1\over x^2+\va^2}\ x ~, \cr 
&a' ~=~ a(x,a)^{-1}\ a ~, ~~ 
\vert a(x,a)\vert ~=~ x^2+\va^2 ~, \cr &\cr 
&k ~=~ k(x,a) ~\equiv ~\pmatrix{ {2\over x^2+\va^2}\ x_i x_j 
- \d_{i j}& 
- {2\va\over x^2+\va^2}\ x_i & 0 \cr &&\cr 
 {2\va\over x^2+\va^2}\ x_j & {x^2-\va^2\over x^2+\va^2} & 0 \cr 
&&\cr 0 & 0 & 1 }}} 
Using \dek$\,$ we note for later use
~$k(x,\id) ~=~ -\ ^t k_x\,$, and also: 
\eqn\klk{k(x,a) ~=~ m(k(x,a))\,k(x,\va)_f 
~=~ r(x)\, k_{{\va\over x^2}\,x} } 
Here we first record (for $x\neq 0$):
\eqn\hlp{\eqalign{ 
  \lim_{\va\to 0}\ x(x,a)  
 ~&=~ - {1\over x^2}\ x ~\equiv~ R x \cr 
  \lim_{\va\to 0}\  a(x,a)  ~&=~ a(x) ~, \qquad 
\vert a(x)\vert ~=~ x^2 ~, \cr 
  \lim_{\va\to 0}\ k(x,a)  ~&=~ r(x) }} 
and \prf{a} becomes: 
\eqn\chkc{\eqalign{ 
\lim_{\va\to 0}\ \va^{{-\D}}\ \Pi^\t_\mu\  \tD^\t(k(x,a))\ 
&\phi (x(x,a),\ \vert a(x,a)\vert^{-1}\va) ~=\cr =&~ 
(x^2)^{{-\D}}\
\hd^\mu(r(x))\ \lim_{\va\to 0}\ \va^{{-\D}}\ \Pi^\t_\mu\ 
 \phi (R x,\ \va) }} 
On the other hand \prff{b} gives (also for $x\neq 0$): 
~$R\tn_x ~=~ \tn_{x''}
 m^{-1}{a''}^{-1} n^{-1}$~ with ~$x'' ~=~ R x$, ~$m ~=~ r(x)$, ~$a'' 
~=~ a(x)$, using the notation introduced in \hlp. Thus,  
\prff{a} also becomes \chkc. 

This finishes the Proof of the intertwining property. On the way 
we have proved also \prja. Indeed, though we have started with 
\prj\ in the generic f-la \prf{a} for the LHS of the intertwining property, 
in the four generating cases above  
we have ~$k_f=\id$, i.e., we could have started with \prja\ in 
\prf{a}.~$\spadesuit$

\medskip

\newsubsec{Boundary-to-bulk intertwining relation} 

\nt
Now we look for the possible operator inverse to ~$L_\chi^\t\,$~ which 
would restore a function on de Sitter space from its boundary 
value, as discussed in \refs{\Wi,\HeSf{--}\DhFr}. Again 
what is new here is that we define it as intertwining operator between 
exactly defined spaces in a more general setting. 
Moreover, we shall construct the operator just from 
the condition that it is an intertwining integral operator. Indeed, 
let us have the operator: 
\eqn\inv{{\tilde L}_\chi^\t 
~:~ C_\chi ~\rra ~ \hc_\chi^\t   ~,}
and try the following Ansatz:
\eqn\inta{ \( {\tilde L}_\chi^\t \ f\) (x,\va) ~=~ 
\int \kc (x,\va;x')\, f(x')\, d x' } 
where ~$\kc (x,\va;x')$~ is a linear operator acting 
from the space $V_\mu$ to the space $U_\t\,$, 
and let us suppose that 
~${\tilde L}_\chi^\t$~ is an intertwining operator, i.e.: 
\eqn\iaac{\htt^\t (g) \circ  {\tilde L}_\chi^\t     
~=~ {\tilde L}_\chi^\t \circ  T^\chi (g) ~, \quad \forall g\in G ~.} 
As in the Theorem above we apply \iaac$\,$ for ~$g=\tn_y,\ha,\hm,R$~ 
and we use the same decompositions as above, so we can present things 
in a short fashion. Applying \iaac$\,$ for ~$g=\tn_y$~ results in the 
fact that ~$\kc$~ depends only on the difference of the $x$ arguments:
\eqn\gga{ \kc (x,\va;x') ~=~ \kc (x-x',\va) ~.} 
Applying \iaac$\,$ for ~$g=\ha$~ results in the 
fact that ~$\kc$~ is homogeneous in its arguments:
\eqn\gga{ \kc (\s x,\s \va) ~=~ \s^{\D-d} \kc (x,\va) ~, \quad \s\in\bbc, 
\ \s\neq 0 } 
Thus, we shall write:
\eqn\ggb{\kc (x,\va) ~=~ \va^{\D-d} \hat\kc \({x\over \va}\) ~, \qquad 
\hat\kc (y) ~\doteq~ \kc (y,1) }
Note now that \gga$\,$ means, in particular, that ~$\kc(0,\va)$~ 
(if it exists) is fixed up to a constant matrix:
\eqn\ggaa{ \kc (0, \va) ~=~ \va^{\D-d} \kc (0,1) ~=~ 
\va^{\D-d} \hat\kc (0) .}
Applying \iaac$\,$ for ~$g=\hm$~ results in the 
fact that ~$\kc$~ has the following covariance property:
\eqn\ggm{ \tD^\t (m)\, \hat\kc (x) ~=~  \hat\kc (m x)\, \hd^\mu (m)  
~,  \quad \forall\, m\in M } 
The above means, in particular, that ~$\Pi^\t_\mu\,\hat\kc(0)$~ 
is $M$-invariant:
\eqn\ggmz{ \hd^\mu (m)\ \Pi^\t_\mu \hat\kc (0) ~=~  
\Pi^\t_\mu\hat\kc (0)\ \hd^\mu (m)  ~,  \quad \forall\, m\in M } 
which then, by Schur's Lemma, means 
that ~$\Pi^\t_\mu \hat\kc (0) ~=~ \s\id_\mu\,$, ~$0\neq\s\in\bbc$. 
Thus, we have: ~$\hat\kc (0) ~=~ \s\, \Pi^\mu_\t\,$, the latter 
being the canonical embedding operator from $V_\mu$ to $U_\t\,$, 
such that ~$\Pi^\t_\mu \circ \Pi^\mu_\t ~=~ \id_\mu\,$.  
Finally, applying \iaac$\,$ for ~$g=R$~ means that:
\eqn\rinv{ \tD^\t (r(x))\ \tD^\t (k_{{\va\over x^2}\,x})\ 
\hat\kc (-{x\over \va} + 
{x^2 +\va^2\over \va y^2}\, y) 
~=~ \( {y^2 \over x^2 +\va^2}\)^{d-\D}\ \hat\kc({x-y\over\va})\  
\hd^\mu (r(y)) }
where we have used the decomposition of ~$k(x,\va)$~ in \klk. 
Now we set ~$y=x$~ and we get: 
\eqn\rinz{ \tD^\t (r(x))\ \tD^\t (k_{{\va\over x^2}\,x})\ 
\hat\kc ({\va\over x^2}\,x) 
~=~ \( {x^2 \over x^2 +\va^2}\)^{d-\D}\ \hat\kc (0)\   
\hd^\mu (r(x)) }
Using \ggm$\,$ for $x=0$ we obtain (using also that $r(x)^2 = \id$): 
\eqn\rinz{ \tD^\t (k_{{\va\over x^2}\,x})\ 
\hat\kc ({\va\over x^2}\,x) 
~=~ \( {x^2 \over x^2 +\va^2}\)^{d-\D}\ \hat\kc (0) }
Next, we make the change ~${\va\over x^2}\,x \to x$~ 
and we get: 
\eqn\ffn{ \hat\kc (x) ~=~ N_\chi^\t\ \( {1 \over x^2 +1}\)^{d-\D}\ 
\tD^\t (k_{-x})\ \Pi^\mu_\t } 
or finally: 
\eqn\fff{ \kc (x,\va) ~=~ N_\chi^\t\ \( {\va \over x^2 +\va^2}\)^{d-\D}\ 
\tD^\t (k_{-{x\over \va}})\ \Pi^\mu_\t }
where ~$N_\chi^\t$~ is arbitrary for the moment and should be fixed 
from the requirement that ~${\tilde L}_\chi^\t$~ is inverse to 
~$L_\chi^\t\,$ (when the latter is true). 

The above operator exists for arbitrary representations 
~$\t$~  of ~$K=SO(d+1)$~ which contain 
the representation ~$\mu$~ of ~$M=SO(d)$. 
We use the standard ~$SO(p)$~ representation 
parametrization: ~$[\ell_1,\dots,\ell_\tp]$, ($\tp\equiv\tpp$),   
where all $\ell_j$ are simultaneously integer or 
half-integer, all are positive except for $p~~even$~ when 
~$\ell_1$~ can also be negative, and they are ordered: 
~$\vert \ell_1\vert\leq \ell_2\leq \dots \leq \ell_\tp$. 
The condition that ~$\t ~=~ [\ell'_1,\dots,\ell'_\thd]$,   
($\thd \equiv \ttd$),  contains ~$\mu ~=~ [\ell_1,\dots,\ell_\td]$, 
($\td \equiv \tdd$), explicitly is:
\eqna\cont
$$\eqalignno{&\vert\ell'_1\vert \leq \ell_1 \leq \dots \leq \ell_\td 
\leq \ell'_\thd ~, \qquad d~~ odd, ~~\thd=\td+1 &\cont a\cr
-&\ell'_1\leq \ell_1 \leq \ell'_1\dots \leq \ell_\td \leq 
\ell'_\thd ~, \qquad d~~ even, ~~\thd=\td &\cont b\cr}$$ 

If one is primarily concerned with the ERs ~$\chi = [\mu,\D]$~ 
it is convenient to chose a 'minimal' representation  
~$\t(\mu)$~ of ~$K=SO(d+1)$~ containing ~$\mu\,$.  
This depends on the parity of ~$d$. Thus, for ~$\mu$~ as above, 
when ~$d$~ is ~{\it odd}~ we would choose: 
\eqn\ppp{ \t(\mu) ~=~ [\ell_1,\ell_1,\dots,\ell_\td] 
\qquad {\rm or}\qquad  \tilde\t(\mu) ~=~ 
[-\ell_1,\ell_1,\dots,\ell_\td] ~, \qquad \mu \cong \tmu ~, } 
while for ~{\it even}~ $d$~ we would choose: 
\eqn\ppp{\t(\mu) ~=~ [\vert\ell_1\vert,\ell_2\dots,\ell_\td] 
~=~ \t(\tilde\mu) ~\cong~ \tilde\t(\mu) ~=~ \tilde\t(\tilde\mu) 
~, \quad \tilde\mu ~=~ [-\ell_1,\ell_2\dots,\ell_\td] ~.}
Thus,  in the odd $d$ case for each ~$\mu$~ we would choose between 
two $K$-irreps which are mirror images of one another, 
while in the even $d$ case to each two mirror-image irreps of $M$ 
we choose one and the same irrep of $K$.

The explicit formulae which appeared until now in the literature are 
actually in the cases in which ~$\t=\t(\mu)$, though there is no such 
interpretation as we have here. In such a restricted 
setting and from other considerations  
formula \fff$\,$ for the scalar case (when both $\mu$ and $\t=\t(\mu)$ are 
scalar irreps) was given by Witten \Wi$\,$,   
while some other nonscalar cases were given in 
\refs{\Wi,\HeSf,\LiTsa\CNSS{--}\MuVib,\Corl\Vol{--}\Yi}.  
Note that in (2.38) of \Wi$\,$ it is written for the conjugated conformal 
weight: ~$\D\to d-\D$, which in our language would mean to work with 
the representation ~$\tch ~=~ [\tmu, d-\D]$~ and to use:
\eqn\ffn{ \kci (x,\va) ~=~ N_\tch^\tt\ \( {\va \over x^2 +\va^2}\)^{\D}\ 
\tD^\tt (k_{-{x\over \va}})\ \Pi^\tmu_\tt }

\medskip 

\newsubsec{Equivalence vs. partial  equivalence} 

\nt 
Either one of the representation equivalences established in 
the previous subsections means that the representations 
~$\hat C_\chi^\tt$~ and ~$C_\chi$~ are\ {\it partially equivalent}. 
In order for them to be\ {\it equivalent}$\,$ it is necessary and 
sufficient that the operators  ~${\tilde L}_\chi^\tt$, $L_\chi^\tt\,$ 
are inverse to each other, i.e., the following relations 
should hold:  
\eqn\inve{L_\chi^\tt \circ {\tilde L}_\chi^\tt ~=~ \id_{C_\chi} ~, 
\qquad {\tilde L}_\chi^\tt \circ L_\chi^\tt ~=~ \id_{\hat C_\chi^\tt} }
If \inve$\,$ do not hold then at least one 
of the representations ~$\hat C_\chi^\tt$~ and ~$C_\chi^\tt$~ 
is topologically reducible. 

For the first relation in \inve$\,$ we have:
\eqn\invc{\eqalign{  \( L_\chi^\tt \circ {\tilde L}_\chi^\tt\ f\) (x) ~=&~  
\lim_{\va\to 0}\ \va^{{-\D}}\ \Pi^\tt_\mu\ 
\( {\tilde L}_\chi^\tt\ f\) (x,\va) ~=\cr 
=&~  \lim_{\va\to 0}\ \va^{{-\D}}\ \Pi^\tt_\mu\ 
\int \kc (x-x',\va)\ f(x')\ d x' }}
 
For the above calculation we interchange the limit and the integration, 
and use the following result from \Wi$\,$ (there for $\D> d/2$): 
\eqn\ddd{ \lim_{\va\to 0}\ \va^{{\D-d}}\ 
\( { \va \over x^2 + \va^2 }\)^{\D}\ ~\sim~ \d(x)} 
substituting for our needs ~$\D\to d-\D$. 

To obtain the proportionality constant in \ddd, and thus fix ~$N_\chi^\tt\,$,  
we first find the Fourier transform of ~$(x^2 + \va^2)^{-\D}$~:
\eqn\fff{ \eqalign{ 
\int {e^{-i p\cdot x}\over (x^2 + \va^2)^{\D} }\ {d x\over (2\pi)^{d/2}} ~=&~ 
{1 \over 2^\D\ \G(\D)}\ 
\int \int_0^\infty d\a\ \a^{\D-1}\  e^{-i p\cdot x -\half\a(x^2 + \va^2)}\  
{d x\over (2\pi)^{d/2}} ~=\cr 
=&~ {1 \over 2^\D\ \G(\D)}\ \int_0^\infty d\a\ \a^{\D- {d\over 2}- 1}
 e^{-\half\({p^2\over \a} + \a\va^2\)} ~=\cr  
=&~ {2 \over 2^\D\ \G(\D)}\ \( {\sqrt{p^2} \over \va}\)^{{\D}-{d\over 2}}\ 
K_{\D-{d\over 2}}(\va\sqrt{p^2}) }}
where we have used f-la 3.47.9 of \RG, involving the 
Bessel function $K_\nu\,$. Note that for ~$\va\to 0$~ 
the above formula goes to formula (5.2) of \DMPPT\ (with ~$\D= h+c$), 
in particular, the RHS of \fff$\,$ goes to: 
\eqn\lll{ {\G({d\over 2}-\D)\over 2^\D\ \G(\D)}\ 
\({p^2\over 2}\)^{\D-{d\over 2}}  ~, \quad {d\over 2}-\D \notin \bbz_-} 
For \lll$\,$ one uses the relation: 
\eqn\kkk{ \lim_{\b\to 0}\ 2\ \b^\nu\ K_\nu(\b) ~=~ 2^\nu\ \G(\nu) 
~, \quad \nu \notin \bbz_- }

Now we can find the necessary limit in \ddd$\,$ (again using \kkk): 
\eqn\mmm{\eqalign{ \lim_{\va\to 0}\ \va^{{\D-d}} 
\( { \va \over x^2 + \va^2 }\)^{\D}\ ~=&~ \lim_{\va\to 0}\ 
{2\ \va^{{2\D-d}}\over 2^\D\ \G(\D)} \int 
\( {\sqrt{p^2} \over \va}\)^{\D-{d\over 2}}  
K_{\D-{d\over 2}}(\va\sqrt{p^2}) {e^{i p\cdot x}\ d p\over (2\pi)^{d/2}} 
~=\cr =&~\lim_{\va\to 0}\ 
{2 \over 2^\D\ \G(\D)} \int 
\(\va\sqrt{p^2} \)^{{\D}-{d\over 2}} 
K_{\D-{d\over 2}}(\va\sqrt{p^2})  {e^{i p\cdot x} d p\over (2\pi)^{d/2}} 
~=\cr =&~ {\G(\D-{d\over 2}) \over 2^{d\over 2}\ \G(\D)}\ \int 
{e^{i p\cdot x}\ d p\over (2\pi)^{d/2}} ~=\cr =&~ 
{\pi^{d\over 2}\ \G(\D-{d\over 2}) \over 
\G(\D)}\ \d(x) }} 
where we have used:
\eqn\dlt{ \d(x) ~=~ \int {e^{i p\cdot x}\ d p\over (2\pi)^{d}} ~. }

In the scalar case 
prompted by \mmm\ (with $\D\to d-\D$) we choose the constant in \fff$\,$ as:
\eqn\ccc{N_\chi^\tt ~=~ {\G(d-\D) \over \pi^{d\over 2}
\G({d\over 2}-\D)} ~, \qquad \D\neq d+k ~, ~~k\in\bbz_+ ~{\rm for}~ d ~odd ~.} 
This is the choice made in \FMMRa\ (for ~$\D\to d-\D$)  
from other considerations. In the general case we choose 
~$N_\chi^\tt$~ as:
\eqn\nrmm{\eqalign{ 
N_\chi^\tt ~=&~ N_0\, {\G(\thd -\D ) \over \G(\had -\D) } \ 
\pl_{k=1}^\td (m_k+\had-\D)  \cr 
&m_k\ \doteq \ \vert \ell_k +k-1 +\d_d\vert \ , ~~k=1,\dots,\td \cr 
& \td \equiv \left[\had\right]\ , \quad \thd \equiv \ttd\ ,\quad   
\d_d ~=~ \had-\td ~=~ \cases{ 0  &~ $d$~ even \cr \half  &~ $d$~ odd }}} 
where ~$N_0$~ is a constant independent of $\D$ having no poles 
or zeroes for any ~$\chi$. Note that for ~$d~~odd$~  ~$N_\chi^\tt$~ 
has poles and thus is not defined in the following cases:
\eqn\exla{ \D ~=~ 
\hadd + p ~, \quad p\in\bbz_+ ~, \qquad p\neq m_k -\half\,, ~~k=1,\dots,\td, 
~~ d~~odd\,, }
(which for the scalar case coincide with the exclusion conditions in 
\ccc). We note also the zero cases: 
\eqn\exlb{ 
N_\chi^\tt ~=~ 0, \quad {\rm for}
~ \cases{ \D ~=~ m_k+\had ~, ~~k=1,\dots,\td, & $d$ ~even
\cr  \D ~=~ m_k+\had ~\neq~ \hadd +p, ~k=1,\dots,\td, ~p\in\bbz_+\,, 
& $d$ ~odd \cr 
\D ~=~ \had +p,  ~p\in\bbz_+\,, & $d$ ~odd }}

Further we need also the following fact which for simplicity 
we write for $d$ odd (or for $G\to G'$): 
\eqn\pqr{\lim_{\va\to 0}\ k_{-{x\over \va}} ~=~ r(x)\, R}  

With the choice \ccc$\,$ (or \nrmm) and using \mmm$\,$ and \pqr\ 
the last line of \invc$\,$ gives 
~$f(x)$~ thus establishing the first relation in \inve$\,$ for 
generic values of ~$\D$ (i.e., when ~$N_\chi^\tt$~ is finite 
and nonzero).

As a ~{\it Corollary}~ we recover the fact \Wi$\,$ 
that for generic values of ~$\D$~ 
we can restore a function on de Sitter space from its boundary 
value on $R^d$. Indeed, suppose we have:
\eqn\rst{ \phi(x,\va) ~=~ \int \kc (x-x',\va) f(x') \,d x' ~, \qquad 
\phi \in\hat C_\chi^\tt ~, ~~ f\in C_\chi } 
Then we have for the boundary value:
\eqn\rss{\eqalign{ \psi_0(x) ~\doteq&~ \(L^\tt_\chi\, \phi\) (x) ~=~  
\lim_{\va\to 0}\ \va^{-\D}\ \Pi^\tt_\mu\ \phi(x,\va) 
~=\cr =&~ \lim_{\va\to 0}\ \va^{-\D}\ \Pi^\tt_\mu\ 
\int \kc (x-x',\va) f(x') \,d x' 
~=~ f(x) }} 
 
Now we can prove the second relation in \inve: 
\eqn\invd{\eqalign{  \( \tilde L_\chi^\tt \circ {L}_\chi^\tt\ \phi \) 
(x,\va) ~=&~  
\int \kc (x-x',\va) \(  {L}_\chi^\tt\ \phi \) (x') \,d x' ~=\cr 
~=&~ \int \kc (x-x',\va) 
\lim_{\vv\to 0}\ \vv^{{-\D}}\ \Pi^\tt_\mu\ \phi (x',\vv) ~=\cr 
=&~  \int \kc (x-x',\va)\ \psi_0 (x') \,d x' ~=~ \phi (x,\va) }}
where in the last line we used \rss. 

\bu~ Thus, we have found that the partially equivalent representations 
~$\hat C_\chi^\tt$~ and ~$C_\chi$~ ($\chi =[\mu,\D]$) are equivalent iff 
~$\D$~ is not in the excluded ranges given in \exla, \exlb.

This result may be used for the conjugate situation ~$\chi\to \tch$.
The constant then is:
\eqn\nrmu{ N_\tch^\tt ~=~ \tilde N_0\, {\G(\D-\td) \over \G(\D-\had)}\ 
\pl_{k=1}^\td (m_k-\had+\D) } 
where ~$\tilde N_0$~ is a constant independent of $\D$ having no poles 
or zeroes for any ~$\chi$. 
The cases when ~$N_\tch^\tt$~ is not defined are: 
\eqn\ecla{\D ~=~ \td - p ~, \quad p\in\bbz_+ ~, 
\qquad p\neq m_k -\half\,, ~~k=1,\dots,\td, 
~~ d~~odd\,, }
while the zero cases are: 
\eqn\eclb{ 
N_\tch^\tt ~=~ 0, \quad {\rm for}
~ \cases{ \D ~=~ \had - m_k ~, ~~k=1,\dots,\td, & $d$ ~even
\cr  \D ~=~ \had-m_k ~\neq~ \hadd +p, ~k=1,\dots,\td, ~p\in\bbz_+\,, 
& $d$ ~odd \cr 
\D ~=~ \had -p,  ~p\in\bbz_+\,, & $d$ ~odd }}

\bu~ Thus, we find  that the partially equivalent 
representations ~$\hat C_\tch^\tt$~ and  ~$C_\tch$~ ($\tch =[\tmu,d-\D]$) 
are equivalent iff 
~$\D$~ is not in the excluded ranges given in \ecla, \eclb. 

\medskip

\newsubsec{Further intertwining relations}

\nt 
We start by recording the second limit of the bulk functions 
(which we mentioned towards the end of Section 4). We take 
as in the previous subsection ~$\psi_0\in C_\chi\,$, 
$\chi =[\mu,\D]$, and ~$\phi\in\hc^\t_\chi$~ expressed 
through $\psi_0$ by  \rst, \rss. We set and calculate: 
\eqn\sndl{ \eqalign{ \phi_0(x) ~\doteq&~ 
\lim_{\va\to 0}\ \va^{\D-d}\ \Pi^\tt_\tmu\  \phi(x,\va) 
~=\cr =&~ \lim_{\va\to 0}\ \va^{\D-d}\ \Pi^\tt_\tmu\ 
\int \kc (x-x',\va) \psi_0(x') \,d x' 
~=\cr  =&~ N_\chi^\tt\ \int {d x'\over (x-x')^{2(d-\D)} }\ 
\Pi^\tt_\tmu\ \tD^\tt (r(x)R)\ \Pi^\mu_\tt\ \psi_0 (x') 
~=\cr  =&~ N_\chi^\tt\ \int {d x'\over (x-x')^{2(d-\D)} }\ 
\hd^\tmu (r(x)\  \psi_0 (x') 
~=\cr  =&~ {N_\chi^\tt\over \g_\tch} \ \int d x'\ G_\tch (x-x')\ 
\psi_0 (x') }}
Since ~$\psi_0\in C_\chi$~ from the intertwining property 
of ~$G_\tch$~ follows that ~$\phi_0\in C_\tch\,$, 
$\tch =[\tmu,d-\D]$. This is valid for generic representations - 
in the exceptional cases it may happen that ~$\phi_0=0$~ 
(when $\psi_0$ is in the kernel of $G_\tch$) or that the 
asymptotic expansion contains logarithms, (for the latter 
cf., e.g., (7.45) of \DMPPT, \Dobp). 

Thus, we have established \dud~ at generic representation points. 
The two different limits of the bulk field ~$\phi$~ are given 
by the coupled fields ~$\phi_0$~ and ~$\psi_0$\ (the latter we 
denote also by $\co$). 

Formulae \sndl~  mean that if we define: 
\eqn\bbb{ \cac_\tch^\tt ~:~ \hc_\chi^\tt ~\rra~ C_\tch } 
with the  action: 
\eqn\aab{ (\cac_\tch^\tt \phi ) (x) ~=~ \lim_{\va\to 0}\ 
\va^{\D-d}\ \Pi^\t_\tmu\ \phi (x, \va) } 
then  we can show as in Theorem the intertwining 
property:
\eqn\intsc{ \cac_\tch^\tt \circ \hat T^\t (g) 
~=~ T^\tch (g) \circ \cac_\tch^\tt 
~, \quad \forall g\in G ,}
since in the Proof of the Theorem we have used only 
the fact of the existence of the  limit. 
On the other hand  ~$\cac_\tch^\tt$~ is equal to  
~$L_\tch^\t$~ because of \dud. 

Next we note that the intertwining property \intsc$\,$ 
is fulfilled if we take the following as a defining relation: 
\eqn\intsa{ {\cac'}_\tch^\tt ~=~ G_\tch \circ L_\chi^\tt }
Expectedly, we get the same result as in \sndl\ (up to 
a multiplicative constant):
\eqn\into{ \eqalign{ \( {\cac'}_\tch\,\phi \) (x) ~=&~ 
  \( G_\tch \circ L_\chi^\tt \,\phi \) (x) ~=\cr 
=&~ \int d x'\, G_\tch (x-x') \( L_\chi^\tt \,\phi \) (x') ~=\cr 
=&~  \int d x'\, G_\tch (x-x')\, \psi_0 (x') }} 

Gathering everything we have in this subsection we obtain 
the following relation between three of the operators 
under consideration:
\eqn\rlr{ L^\tt_\tch ~=~  {N_\chi^\t\over \g_\tch}\ 
G_\tch \circ L_\chi^\t }

At generic points from this we can obtain a lot of 
interesting relations, e.g., applying ~$\tilde L^\t_\chi$~ 
from the right we get:
\eqn\sss{  L^\t_\tch \circ \tilde L^\t_\chi  ~=~  
{N_\chi^\tt\over \g_\tch}\ G_\tch } 
which is in fact \sndl. Applying ~$G_\chi$~ from the left 
we get:
\eqn\ssa{ L^\t_\chi ~=~ {\g_\tch \over  N_\chi^\t}\ 
G_\chi \circ   L^\tt_\tch } 
and making the change ~$\chi\llra\tch$~ in the last equality 
we get:
\eqn\ssb{ L^\tt_\tch ~=~ {\g_\chi \over  N_\tch^\tt}\ 
G_\tch \circ   L^\t_\chi } 

Comparing \rlr, \ssb, we obtain the following relation  
between the normalization constants:
\eqn\nnn{ N_\chi^\t\ N_\tch^\tt ~=~ \g_\chi\ \g_\tch ~=~ C\, \rho (\chi)} 
where ~$\rho (\chi)$~ is the analytic continuation of the 
Plancherel measure for the Plancherel formula 
contribution of the principal series of unitary irreps of ~$G$, 
and the last equality was shown in \KnSt. (The constant $C$ is independent 
of ~$\chi\,$.) 
The Plancherel measure itself is given as follows \Nai, \Hir:\foot{The 
principal series is obtained for ~$c$~ pure imaginary.} 
\eqna\pla
$$\eqalignno{\rho (\chi) ~=~ \rho (\tch) ~=&~ \rho_0 (\mu)\  
\( \prod_{1\leq i<j\leq \td} (m_j^2-m_i^2) \)\ 
{\G(\d_d +c)\,\G(\d_d -c) \over  \G(c)\,\G(-c)}\   
\prod_{k=1}^\td (m_k^2 - c^2) 
\qquad  
~&\pla {}\cr &&\cr &&\cr 
&c ~\equiv~  \D -\had\ , \qquad 
\rho_0 (\mu) ~=~ \cases{ 1 ~~&~~ $d$~ even \cr &\cr 
\pl_{k=1}^\td m_k ~~&~~ $d$~ odd } }$$
(cf. \nrmm). 
In the scalar case, when ~$\ell_j =0$~ for all $j$, we have:
\eqn\scsc{ \rho (\chi) ~=~ 
C'~ {\G(\had +c)\, \G(\had-c) \over \G(c)\,\G(-c) }
~=~ C'~ {\G(\D)\ \G(d-\D) \over \G(\D-{d\over 2})\ \G({d\over 2}-\D)}
}
the constant $C'$ being independent of ~$\D$. 

Another consequence of \nnn~ is that the constants ~$N^\t_\chi$~ 
do not depend on ~$\t$~ since the constants ~$\g_\chi$~ do not. 
Thus, we drop their superscript ~$\tau\,$. 

The above relation between these constants may be 
obtained also if we use the reconstruction of the 
field ~$\phi$~ from ~$\phi_0$~:
\eqn\rcc{ \phi (x,\va) ~=~ \int \kci (x-x',\va)\, \phi_0 (x') d x' 
~=~ \( \tilde L^\t_\tch\ \phi_0\) (x,\va) } 
Proceeding as in \sndl~ we have:
\eqn\snld{\eqalign{ 
 \psi_0 (x) ~=&~ (L^\t_\chi\, \phi) (x) ~=~ 
{N_\tch \over \g_\chi } \int d x'\ G_\chi (x-x')\, \phi_0 (x') ~=\cr 
=&~ {N_\chi\, N_\tch \over \g_\chi\,\g_\tch } 
\int d x'\ d x''\ G_\chi (x-x')\ G_\tch (x'-x'')\  \psi_0 (x'') ~=\cr &\cr 
=&~ {N_\chi\, N_\tch \over \g_\chi\,\g_\tch }\ \psi_0 (x)}} 
where we have substituted \sndl~ in the last but one line, 
and used \nrm~ in the last line.

Another interesting relation includes a convolution of a 
~$K$-kernel and a ~$G$-kernel. For this we apply 
~$\tilde L^\tt_\tch$~ to \sss$\,$ from the left and 
changing  ~$\chi\llra\tch$~ we get (denoting $x_{ij} \equiv x_i -x_j$):
\eqn\ssc{ \tilde L^\t_\chi \circ\, G_\chi ~=~ {\g_\chi \over  N_\tch} 
\ \tilde L^\tt_\tch }
which applied to ~$f\in C_\tch$~ gives:
\eqna\intos
$$\eqalignno{ ( \tilde L_\chi^\t \circ & G_\chi \, f ) 
(x_1,\va)  ~=~ \int d x_2\, \kc (x_{12},\va) \( G_\chi \, f \) (x_2) ~=\cr 
~=&~ N_\chi\,\g_\chi\ \int d x_2\, d x_3\,  
\( {\va\over x_{12}^2 +\va^2}\)^{d-\D}\ \( {1\over x_{23}^2}\)^{\D}\  
\tD^\t (k_{-{x_{12}\over \va}})\ \Pi^\mu_\tt\, \hd^\mu (r(x_{23}))\ 
f (x_3) ~=\cr =&~ N_\chi\,\g_\chi\ \int d x_2\, d x_3\,  
\( {\va\over x_{12}^2 +\va^2}\)^{d-\D}\ \( {1\over x_{23}^2}\)^{\D}\  
\tD^\t \(k_{-{x_{12}\over \va}}\, r(x_{23})\)\ 
\Pi^\tmu_\tt\ f (x_3)  ~=\cr =&~ 
\g_\chi\ \int d x_3\, \( {\va\over x_{13}^2 +\va^2}\)^{\D}\ 
\tD^\t \(k_{-{x_{13}\over \va}}\)\ 
\Pi^\tmu_\tt\ f (x_3) ~=\cr =&~ 
{\g_\chi\over N_\tch}\ \int d x_3\,
\kci (x_{13},\va)\ f (x_3) ~=~ {\g_\chi\over N_\tch}\ 
\( \tilde L^\tt_\tch\ f\) (x_1,\va)  
&\intos{} }$$ 

Finally, we notice that all operators that we have used 
may be found on  the following commutative diagram:
\eqn\dgr{ 
\matrix{&&&&\cr 
&&\hc_\tch^\t=\hc_\chi^\t\ [\phi]&&\cr 
&&&&\cr &&&&\cr 
&L_\tch^\tt\ \swarrow\nearrow 
\tilde L_\tch^\tt       
&& \tilde L_\chi^\t\     
\nwarrow\searrow
L_\chi^\t &  \cr
&&&&\cr &&&&\cr 
&& G_\tch&& \cr
C_\tch\,[\phi_0] &&{\lla\atop\rra} && C_\chi\,[\co,\psi_0]\cr 
&& G_\chi&& \cr  } }

\vskip 5mm \newsec{Comments and outlook} 

\nt 
From the point of view of calculating the conformal 
correlators \refs{\Wi,\HeSf{--}\DhFr} of the conformal field 
~$\co~\in C_\chi\,$, 
($\chi =[\mu,\D]$, ~$\D> {d\over 2}$)~ from the de Sitter configuration 
described by the field ~$\phi~\in\hc_\chi^\t\,$,  
the  conditions \ecla, \ecla{} are sufficient. 
If one needs to consider some of these exceptional ~$\D$-values 
one has to take another choice for ~$N_\tch$~ (e.g., like 
the one in \FMMRa$\,$ for ~$\D= {d\over 2}$), and thus obtain a nonzero 
operator ~$\tilde L^\t_{\tch}\,$. But also then 
~$\tilde L^\t_{\tch}$~ will not be inverse to 
~$L^\t_{\tch}\,$. Of course, one should 
repeat separately the calculations starting from the 
analogues of \fff. Then one may conclude that 
the asymptotic behaviour of ~$\phi$~ is logarithmic, 
cf., e.g., \refs{\FMMRa,\CKA}. We do not discuss this anymore here, 
but we should note that on the conformal side 
logarithmic behaviour occurs also when one is 
renormalizing intertwining operators acting 
between reducible ERs, cf., e.g., \DMPPT, \DP.

{}From the point of view of the direct correspondence 
between  the conformal field ~$\co~\in C_\chi\,$, and 
the bulk field ~$\phi~\in\tc_\chi^\t\,$~  
there are a lot of excluded points in the region directly 
interesting for the applications ~$\D>  {d\over 2}$, cf. \exla, \exlb.  
In particular, this  means that some conformal representations can not be 
extended from the boundary to the bulk, cf. \FeFra, \FeFrb, for 
discussion and earlier references. We do not consider this 
further here since this question should be discussed 
taking into account the subrepresentation structure of 
the reducible representations occurring at the excluded points, 
cf. \Dobp. 

Naturally, one should take into account also the supersymmetry 
aspects of the correspondence. There was a lot of work on this 
already, e.g., \refs{\NiSe,\GNW,\GuMi,\KRN,\GuMa,
\FFZ,\OzTe\AnFea{--}\FLZ,\GMZa,\FeZa\AnFeb\OlRd\FPZ\GMZb{--}\AnFec}. 
However, in order to implement our approach we shall need also 
results from \refs{\DPm\DPu\DPp{--}\DPf} 
on the representations of the conformal superalgebra 
~$su(2,2/N)$~ and  supergroup ~$SU(2,2/N)$. 
In particular, we shall use the classification of 
all positive energy unitary irreducible representations of ~$su(2,2/N)$~  
(including explicit parametrization in terms of representation 
characteristics) \DPu, \DPp, the explicit parametrization of the 
superfield content of the massless UIRs of $su(2,2/N)$ \DPu, 
the elementary representations of $SU(2,2/N)$ and 
intertwining operators between ERs \DPf. 

\vskip 10mm

\noindent 
{\bf Acknowledgments.} ~~The author is thankful to the 
Deutsche Forschungsgemeinschaft (DFG) for the financial 
support as Guest Professor at ASI (TU Clausthal) and to the 
Director of ASI ~Prof. Dr. H.-D. Doebner for the hospitality. 
The author was partially supported by BNFR under contract $\Phi$-643.

\baselineskip=10pt
\parskip 0pt 
\listrefs

\np \end